\documentclass[12pt]{article} \usepackage{amsmath, amssymb, amsfonts,
amstext, amsbsy, array, graphicx} \usepackage{float, epic, epsfig,
latexsym, curves, wrapfig, subfigure, graphics}

\newwrite\ffile\global\newcount\figno \global\figno=1

\def\writedef#1{}

\input epsf \def\figin{\epsfcheck\figin}\def\figins{\epsfcheck\figins}
\def\epsfcheck{\ifx\epsfbox\UnDeFiNeD \message{(NO epsf.tex, FIGURES
WILL BE IGNORED)}
\gdef\figin##1{\vskip2in}\gdef\figins##1{\hskip.5in}
instead \else\message{(FIGURES WILL BE INCLUDED)}%
\gdef\figin##1{##1}\gdef\figins##1{##1}\fi}

\def\figinsert{} \def\ifig#1#2#3{\xdef#1{fig.~\the\figno}
\writedef{#1\leftbracket fig.\noexpand~\the\figno}%
\figinsert\figin{\centerline{#3}}\medskip\centerline{\vbox{\baselineskip12pt
\advance\hsize by -1truein\center\footnotesize{ Fig.~\the\figno.} #2}}
\bigskip\endinsert\global\advance\figno by1}
\def\endinsert{}

\newcommand{\spinl}{l}
\newcommand{\Nsusy}{\mathcal{N}}
\newcommand{\Afluct}{\mathcal{A}}

\newcommand{\tHooft}{'t~Hooft}
\def\N{{\cal N}}

\DeclareMathOperator{\tr}{tr}

\DeclareMathOperator{\diag}{diag}

\begin{document}

\baselineskip 18pt 
\def\beq{\begin{equation}} \def\eeq{\end{equation}}
\newcommand{\bdm}{\[} \newcommand{\edm}{\]}
\newcommand{\bea}{\begin{eqnarray}}
\newcommand{\eea}[1]{\label{#1}\end{eqnarray}}
\renewcommand{\Re}{\mbox{Re}\,} \renewcommand{\Im}{\mbox{Im}\,}
\renewcommand{\a}{\alpha} \def\tm{\theta_-} \def\tp{\theta_+}
\def\fp{\phi_+} \def\fm{\phi_-} \def\dtq{(\partial_r\theta_-)^2}
\def\ptq{(\partial_\alpha\theta_-)^2} \def\dt{\partial_r\theta_-}
\def\pt{\partial_\alpha\theta_-} \def\N{{\cal N}}

\renewcommand\footnoterule{\vspace*{-3pt}
\hrule width 3in height 0.4pt \vspace*{5.6pt}}


\thispagestyle{empty}

\renewcommand{\thefootnote}{\fnsymbol{footnote}}

$\phantom{+}$ \hfill { MPP-2006-3 }

\bigskip\bigskip\bigskip

\begin{center} \noindent \Large \bf
Instantons on D7 brane probes and \\[2mm] 
 AdS/CFT with flavour
\end{center}

\bigskip\bigskip\bigskip

\centerline{ \normalsize \bf R.~Apreda $\negthinspace {}^a$, 
J.~Erdmenger $\negthinspace {}^a$ 
 \footnote[1]{\noindent \tt  
jke@mppmu.mpg.de}, N.~Evans $\negthinspace {}^b$,
J.~Gro\ss e $\negthinspace {}^a$ and Z.~Guralnik $\negthinspace {}^c$}

\medskip\bigskip\bigskip

\bigskip

\centerline{ $^a$ {\it Max Planck-Institut f\"ur Physik (Werner
Heisenberg-Institut)} } \centerline{\it F\"ohringer Ring 6, D - 80805
M\"unchen, Germany}

\bigskip\bigskip

\centerline{ $^b$ {\it Department of Physics, Southampton
University}} \centerline{\it Southampton SO17 1BJ, United Kingdom}

\bigskip\bigskip

\centerline{ $^c$ {\it Massachusetts Institute of Technology,
    Cambridge MA 02139, USA}}

\bigskip
\bigskip\bigskip\bigskip

\renewcommand{\thefootnote}{\arabic{footnote}}

\centerline{\bf \small Abstract}

\medskip

Recent work on adding flavour to the generalized AdS/CFT correspondence is
reviewed. In particular, we consider instanton configurations on two
coincident D7 brane probes. These are matched to the Higgs branch of the dual
field theory. In $AdS_5\times S^5$, the instanton generates a flow of
the meson spectrum. For non-supersymmetric gravity backgrounds, the
Higgs branch is lifted by a potential, which has non-trivial physical
implications. In particular these configurations provide a gravity dual
description of Bose-Einstein condensation and of a thermal phase
transition. Based on talk given by J.~Erdmenger
at the  RTN Workshop 
``Constituents, Fundamental Forces and Symmetries of the Universe'', Corfu,
Greece, 20-26th September 2005.

\medskip

{\small \noindent }

\newpage

\section{Introduction}

D7 brane probes have proved a versatile tool for including quark
fields into the AdS/CFT correspondence. Strings
stretching between the D7s and the D3 branes of the original
AdS/CFT construction provide ${\cal N}=2$ fundamental
hypermultiplets \cite{add1}. Karch and Katz \cite{KK} proposed
that the open string sector on the world-volume of a probe D7
brane is holographically dual to quark--anti-quark bilinears $\bar
\psi \psi$. There have been many studies using probe D7s in a
variety of gravity backgrounds \cite{Myers1,d3d7}.  In this way a
number of non-supersymmetric geometries have been shown to induce
chiral symmetry breaking \cite{BEEGK,csb} (related analyses are
\cite{morecsb,add3}), with the symmetry breaking geometrically
displayed by the D7 brane's bending to break an explicit symmetry
of the space. Meson spectra are also calculable \cite{spectra}.

In scenarios involving two or more D7 probes,
the Higgs branch spanned by squark vevs $\langle \bar q q
\rangle$ can be identified with instanton
configurations on the D7 world-volume \cite{Zach,Johannes}.
These configurations are the
standard four-dimensional instanton solutions living in the four
directions of the D7 world-volume transverse to the D3 branes.
The scalar Higgs vev in the field theory is identified with the
instanton size on the supergravity side.
In the case of a probe in AdS space, there is a moduli space for
the magnitude of the instanton size or the scalar vev. In \cite{Johannes}, the
meson spectrum associated with a particular fluctuation about the instanton
background is calculated. The spectrum exhibits a non-trivial spectral flow.

In less
supersymmetric gravity backgrounds, the moduli space is
expected to be lifted by a potential. This potential
may have either a stable vacuum selecting a particular scalar vev, or 
a  run-away behaviour. In  \cite{AEEG},
the Higgs branch of the ${\cal N}=4$ gauge theory at
finite temperature and density is analyzed. For the finite temperature case we
find a stable minimum for the squark vev which undergoes a first
order phase transition as a function of the temperature (or
equivalently of the quark mass). On the other hand, in the
presence of a chemical potential, the squark potential leads to an
instability indicating Bose-Einstein condensation.  

Moreover the potential obtained from evaluating the D7 probe action on a
static instanton configuration may be used to obtain  
information about some aspects of the
stability of brane embeddings into non-supersymmetric gravity backgrounds.
For instance the dilaton-flow 
background of Constable and Myers \cite{Constable}, which
has been 
used to obtain a gravity dual of chiral symmetry breaking in \cite{BEEGK},
is expected to be unstable. However we show that the scalar quark potential
for the brane embedding into this 
background is well-behaved and drives the vev to zero \cite{AEE}.

\section{Higgs branch AdS/CFT dictionary}

Consider a probe of two coincident D7 branes in $AdS_5 \times S^5$. 
This corresponds to two fundamental hypermultiplets in the dual $\N=2$ gauge
theory. The metric of $AdS_5 \times S^5$ is given by
\begin{gather}\label{AdS}
    ds^2 = {H^{-1/2}(r) \eta_{\mu\nu}dx^\mu dx^\nu +%
           H^{1/2}(r) ( d\vec{y}^{\,2} + d\vec{z}^{\,2} ) \, ,} \\
    H(r) = \frac{L^4}{r^4},  \quad    r^2  = \vec{y}^{\,2} + \vec{z}^{\,2},
   \quad  L^4 = 4\pi g_s N_c (\alpha')^2,  \quad  \vec{y}^{\,2} = 
\sum_{m=4}^{7} y^m y^m, \nonumber\\
    C^{(4)}_{0123} = H^{-1} ,  \quad   \vec{z}^{\,2} = (z^8)^2+(z^9)^2, 
 \quad  e^{\phi}  =e^{\phi_\infty}=g_s.  \nonumber
\end{gather}
Two D7-branes are
embedded into this geometry according to
$  z^8 = 0$,   $z^9 = (2\pi\alpha')m $.
This
leads to the induced metric
\begin{gather}\label{D7geom}
\begin{split}
  ds^2_{\text{D7}} =
     H^{-1/2}(r) \eta_{\mu\nu}dx^\mu dx^\nu +
     H^{1/2}(r) d\vec{y}^{\,2}, 
 \quad  r^2 = y^2 +  (2\pi\alpha')^2 m^2, \; \; y^2 \equiv y^m y^m.
\end{split}
\end{gather}
The parameter $m$ corresponds to the mass of the fundamental
hypermultiplets in the dual $\Nsusy=2$ theory. 

The effective action describing D7-branes in a curved background
is
\begin{align}\label{ac}
S &= T_7 \int  \sum_r C^{(r)} \wedge \tr e^{2\pi  \alpha'
F}
\, + T_7\int\,d^{8}\xi\, \sqrt{g} \, \frac{(2\pi
\alpha')^2}{2} \, \tr \left( F_{\alpha\beta}F^{\alpha\beta}
\right) + \cdots \, , 
\end{align} where we have not written terms involving fermions and scalars.
This action is the sum of a Wess-Zumino term, a Yang-Mills term,
and an infinite number of corrections at higher orders in
$\alpha'$ indicated by $\cdots$ in \eqref{ac}. Since we need to
consider at least two flavors (two D7's) in order to have a Higgs
branch, the DBI action is non-Abelian. The correspondence between
instantons and the Higgs branch suggests that the equations of
motion should be solved by field strengths which are self-dual
with respect to a flat four-dimensional metric.  We
work to leading order only in the large \tHooft{} coupling
expansion generated by AdS/CFT duality, which allows one to only
consider the leading term in the $\alpha'$ expansion of the
action.  Constraints on unknown higher order terms arising from
the existence of instanton solutions, as well as the exactly known
metric on the Higgs branch, are discussed in
\cite{Zach}.

At leading order in $\alpha'$, field strengths which are self dual
with respect to the flat four-dimensional metric $ds^2 =
\sum_{m=4}^7 dy^mdy^m$ solve the equations of motion, due to a
conspiracy between the Wess-Zumino and Yang-Mills term. Inserting
the explicit AdS background values \eqref{AdS} for the metric and
Ramond-Ramond four-form into the action for D7-branes embedded
as given below \eqref{AdS},  with non-trivial field strengths only
in the directions $y^m$,  gives
\begin{align}\label{theactn}
\begin{split}
  S &= \frac{T_7 (2\pi\alpha')^2}{4} \int\, d^4x\,d^4y\, H(r)^{-1} \,
       \left(-\frac{1}{2}\epsilon_{mnrs}F_{mn}F_{rs} +
       F_{mn}F_{mn}\right)  \\
    &= \frac{T_7 (2\pi\alpha')^2}{2} \int \, d^4x\,d^4y\,  H(r)^{-1} F_-^2 \, ,
\end{split}
\end{align}
where $F^-_{mn} =
\frac{1}{2}(F_{mn}-\frac{1}{2}\epsilon_{mnrs}F_{rs})$. Field
strengths $F^-_{mn} = 0$, which are self-dual with respect to the
flat metric $dy^mdy^m$, manifestly solve the equations of motion.
These solutions correspond to points on the Higgs branch of the
dual $\Nsusy=2$ theory.  Strictly speaking, these are points on a
mixed Coulomb-Higgs branch if $m \ne 0$ (for details see \cite{Johannes}). 
We emphasize that in order
to neglect the back-reaction due to dissolved D3-branes, we are
considering a portion of the moduli space for which the instanton
number $k$ is fixed in the large $N_c$ limit.

For $m = 0$, the AdS geometry \eqref{AdS} together
with the embedding \eqref{D7geom},  is invariant
under $SO(2,4) \times SU(2)_L \times SU(2)_R \times U(1)_R \times
SU(2)_f$. The combination $SU(2)_L \times SU(2)_R$ acts as $SO(4)$
rotations of the coordinates $y^m$.  The $SO(2,4)$ factor is the
conformal symmetry of the dual gauge theory. The $SU(2)_L$ factor
corresponds to a global symmetry of the dual gauge theory, while
$SU(2)_R \times U(1)_R$ corresponds to the R symmetries. Finally
$SU(2)_f$ is the gauge symmetry of the two coincident D7-branes
which, at the AdS boundary, corresponds to the flavor symmetry of
the dual gauge theory.

For $m\ne 0$,  the symmetry is broken to $SO(1,3) \times SU(2)_L
\times SU(2)_R \times SU(2)_f$.  This is broken further if there
is an instanton background on the D7-branes. We  focus on that
part of the Higgs branch, or mixed Coulomb-Higgs branch, which is
dual to a single instanton centered at the origin $y^m=0$. The
instanton, in ``singular gauge,'' is given by
\begin{align}\label{thinst}
A_\mu = 0,\qquad A_m = \frac{2Q^2
\bar\sigma_{nm}y_n}{y^2(y^2 + Q^2)} \, ,
\end{align}
where $Q$ is the instanton size and $y_m$ denote the four coordinate
directions parallel to the D7 branes but perpendicular to the D3
branes. Moreover 
$\bar\sigma_{mn} \equiv \frac{1}{4}(\bar\sigma_m \sigma_n - 
\bar\sigma_n \sigma_m)$,
\  $\sigma_{mn}   \equiv \frac{1}{4}(\sigma_m \bar\sigma_n - 
\sigma_n \bar\sigma_m)$,
\  $\sigma_m \equiv (i\vec\tau, 1_{2\times 2})$, 
with $\vec\tau$ being the three Pauli-matrices. We choose singular
gauge, as opposed to the regular gauge in which $A_n = 2
\sigma_{mn}y^m/(y^2 + Q^2)$,  because of the improved
asymptotic behaviour at large $y$. In the AdS setting,  the Higgs
branch should correspond to a normalizable deformation of the
background at the origin of the moduli space.  The singularity of
\eqref{thinst} at $y^m =0$ is not problematic for computations of
physical (gauge invariant) quantities.
The instanton \eqref{thinst} breaks the symmetries to
$ SO(1,3) \times SU(2)_L \times \diag (SU(2)_R
\times SU(2)_f)\, $ and corresponds to a point on the
Higgs branch
$q_{i \alpha }  =  \, v \, \varepsilon_{i \alpha
} $,  $v = \frac{Q}{2\pi \alpha'}$,
where $q_{i \alpha  }$ are scalar components of the fundamental
hypermultiplets, labeled by a $SU(2)_f$ index $i=1,2$, and a
$SU(2)_R$ index $\alpha =1,2$.  All the broken symmetries are
restored in the ultraviolet (large $r$), where the theory becomes
conformal.

\section{Fluctuations and Spectral Flow}

The simplest (non-Abelian) ansatz for fluctuations ${\cal
A}_\mu$ about the instanton background is given by
\begin{gather} \label{ansatz}
  \Afluct_\mu{}^{(a)} = \xi_\mu(k) f(y) e^{ik_\mu x_\mu} \tau^a \, , \quad y^2
  \equiv y^m y^m \, . 
  \end{gather}
Greek indices lable the
four Minkowski directions. (\ref{ansatz}) 
is a singlet under $SU(2)_L$ and a triplet under 
$\diag(SU(2)_R\times SU(2)_f)$. The equation of motion for these fluctuations
becomes 
\begin{align}
  0 = \biggl[ \frac{M^2 R^4}{(y^2 + (2\pi\alpha')^2m^2)^2}
                - \frac{8 Q^4}{y^2(\rho^2+Q^2)^2}
                + \frac{1}{y^3}\partial_y (y^3 \partial_y
                )
 \biggr] f(y) \, ,  
                \label{eqn:ansatzeom}
\end{align}
where $M^2 = -k_\mu k_\mu$.  To determine the spectrum,  we 
find the values of $M^2$ for which this equation admits
normalizable solutions. The spectrum is plotted in Figure 1. 

\begin{center}
  \includegraphics[width=0.5\linewidth]{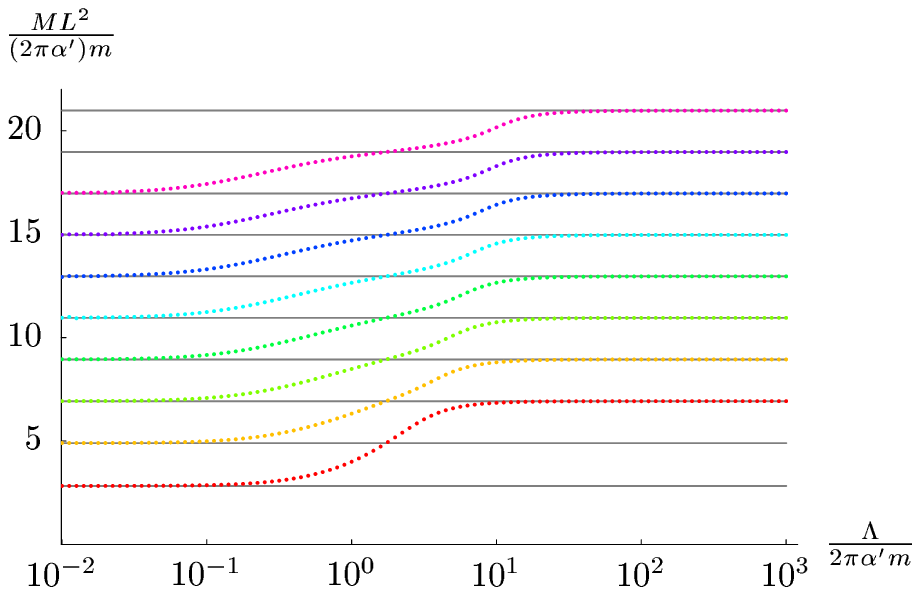}
  {
  \begin{tabular}{r@{ }p{0.45\linewidth}} \small Figure 1. 
    & \small Meson masses as function of the
         Higgs VEV. 
  \end{tabular}
  }\\[1ex]
\end{center}
%

When flowing from
zero to infinite instanton size, the meson 
spectrum $M(n,l)$ is shifted to a 
higher spherical harmonic on $S^3$ with $l=2$. 
It is straightforward to explain this shift.
In the limits of zero or infinite instanton size, we have
\begin{align}
  0 = \biggl[ \frac{{\tilde M}^2}{({\tilde y}^2 + 1)^2}
                - \frac{\spinl (\spinl +2 )}{{\tilde y}^2}
                + \frac{1}{{\tilde y}^3}\partial_{\tilde y}
                ({\tilde y}^3 \partial_{\tilde y} ) \biggr]
  f({\tilde y}), \quad \tilde y \equiv \frac{y} {2 \pi \alpha' m} \, ,
 \label{asy}
\end{align} with  $l = 0,2$. 
This is a special case of the
equations found in \cite{Myers1} for fluctuations
 about the trivial background without any instantons
of the form
\begin{align}
{\cal A}^{\mu} = \xi^{\mu}(k)e^{ik_{\mu}x^{\mu}}f(y) {\cal
Y}_l(S^3) \, .
\end{align}
Here
${\cal Y}_\spinl$ are spherical harmonics on $S^3$ 
corresponding to the
 $(l/2,l/2)$ representation of $SU(2)_L \times
SU(2)_R$. 
In \cite{Myers1} it was found that the spectrum is given by
\begin{align}
  {\tilde M}^2 & = 4(n+\spinl+1)(n+\spinl+2)\, .  \qquad
\end{align}
In our case, at infinite instanton size in singular gauge, the instanton is
given by
\begin{align}
A_n = 2 \frac{\bar\sigma_{mn}y^m}{y^2} \, .
\end{align} 
This instanton may be removed by a gauge transformation of the form
$U = \sigma^m y^m/|y|\, $ which gives $A_n=0$.
In this gauge the fluctuations become
\begin{equation}
{\cal A}_\mu{}^{(a)} = \xi_\mu(k) f(y)
 e^{ik_\mu x_\mu} \frac{y^m  y^n}{y^2} \sigma^m\tau^a\bar\sigma^n \, .
\end{equation}
Here
$\sigma^m\tau^a\bar\sigma^n $ corresponds exactly to the 
$\spinl=2$ spherical harmonic. This explains the shift in spectrum.

\section{Higgs potential for non-supersymmetric backgrounds}

\subsection{Chemical potential}

As the simplest example of a potential generated on the Higgs
branch, we first consider \cite{AEEG} the case of  finite chemical
potential and zero temperature.    
We consider a nonzero chemical
potential for the isospin. 
We allow a spurious gauge
field associated with the $\tau^3$ component of isospin
to acquire a VEV, $\mu$, in its $A^0$ component. This includes generic
fermion and scalar Lagrangian terms for fields with isospin charge
$e$ of the form
\begin{equation} \delta {\cal L} = - \mu e \bar{\psi} \tau^3 \gamma^0 \psi
+ \mu^2 e^2 |\phi|^2 \, .\end{equation}
The first term is a source for the fermionic isospin number density.  In
the path integral, this term places  the theory at finite density. 
The second term
is an unbounded scalar potential which renders the theory
unstable, such that Bose-Einstein condensation is expected. 

This is described in the dual gravity picture as follows.
We add a background $A^0$ in the fixed
instanton background,
\begin{align}\label{backg}A^0 =
\begin{pmatrix} \mu & 0 \cr 0 & -\mu \end{pmatrix}, \qquad A_n =
A_n^{\rm instanton} \, .
\end{align}
On the slice of the Higgs branch corresponding to the single
instanton configurations \eqref{thinst} with modulus $Q$, the
effective potential
at quadratic order in $\mu$ can be determined by inserting
\eqref{backg}
into the D7-brane action. Since the instanton configuration is static
we have $\int d^4 x \; V(Q) = -S_{D7}$,
which gives
\begin{eqnarray}\label{strx} V(Q) =
 T_7 \frac{(2\pi \alpha')^2}{g_s}
\int \, d^4 y\, {\rm tr}\left( \frac{1}{2}\frac{(y^2 +
m^2)^2}{ R^4} F^-_{mn}F^-_{mn} + 2F_{m \mu}F_{m \nu}\eta^{\mu\nu}
\right.\\ \left.
+
\frac{R^4}{(y^2 +
m^2)^2}F_{\mu\nu}F_{\alpha\beta}\eta^{\mu\alpha}\eta^{\nu\beta}\right)\,
, \nonumber
\end{eqnarray} with $y^2 = y^m y^m$. We have split the action into the
pieces involving $F$ in the $x$ and $y$ directions, indicated by
Greek and Roman indices respectively, as well as mixed terms. For
the background \eqref{backg},  the only non-zero contribution to
the potential comes from the mixed term ${\rm tr}\, F_{\mu
m}F_{\nu m}\eta^{\mu\nu} = -{\rm tr}\, [A_0, A_n]^2$, giving
\begin{align}\label{chempot}
V(Q) = -
T_7 \frac{2(4\pi\alpha')^2}{g_s}\mu^2\int d^4y \frac{
Q^4}{y^2(y^2+Q^2)^2} = -
T_7 \frac{2(4\pi^2 \alpha')^2}{g_s}
\mu^2Q^2 \, .
\end{align}
This potential displays an instability which may be interpreted as
Bose-Einstein condensation in the dual field theory.

\subsection{Thermal phase transition} 

The gravitational dual of ${\cal N}=4$ gauge theory at large 't
Hooft coupling and finite temperature is given by 
the AdS-Schwarzschild black-hole
background \cite{Witten:1998zw}. The latter belongs to a general
class of supergravity solutions which, in a choice of coordinates
convenient for our purposes, have the form
 \begin{align}
 \label{bhgeneral}
ds^2 &= f(r)(d\vec x^2 + g(r)d\tau^2)+
h(r)(\sum_{m=4}^7 dy^m dy^m + \sum_{i=8}^9 dZ^i dZ^i),\nonumber\\
e^{-\Phi}&=\phi(r), \qquad r^2 = y^m y^m + Z^i Z^i \, , \nonumber \\[2.5mm]
F^{(5)} &= 4 R^4 (V_{S^5} + ^*V_{S^5})=dC_{(4)}, \qquad
C_{(4)}|_{0123} = s(r)\, dx^0\wedge
dx^1\wedge dx^2\wedge dx^3 \, ,   
\end{align}
For the AdS-Schwarzschild solution, we have \begin{align}\label{AdSSch} f(r)
= \frac{4r^4 + b^4}{4r^2R^2},\,\,\,\,  g(r) =
\left(\frac{4r^4-b^4}{4r^4+b^4}\right)^2,\,\,\,\, 
 h(r) =
\frac{R^2}{r^2},\,\,\,\, s(r) = \frac{r^4}{R^4}\left(1+
\frac{b^8}{16r^8}\right)\, . \end{align} The coordinates $\vec x$ are
the spatial coordinates of the dual gauge theory and $\tau$ is the
Euclidean time direction, which is compactified on a circle of
radius $b^{-1}$, corresponding to the inverse temperature. Note
that the temperature $T\sim b$ only enters to the fourth power.
The D7 embedding in this background  is given by
$Z^9=0 \,$, $Z^8 = z(y)$.

The potential generated on the Higgs branch was calculated in
 \cite{AEEG}. 
Specifically, the action is evaluated on the space of field
strengths which are self-dual\footnote{There are couplings between
world-volume scalars and field strengths at higher orders in
$\alpha'$ which could alter the embedding. However we only
consider the leading term in a large 't Hooft coupling expansion
for which these couplings can be neglected.
} with respect to the induced metric in the directions transverse
to $\tau,\vec x$;
\begin{align}\label{thepot}
V = \frac{T_7 ( 2 \pi \alpha')^2 }{2}\left( \frac{1}{g_s} \int d^4y\,
C^{(4)}_{0123}\, \epsilon_{mnrs}\,{\rm tr}\, F_{mn}F_{rs}
-\frac{1}{2} \int\, d^4y\, \sqrt{-{\rm det}G}\,{\rm
tr}\,F^{mn}F_{mn}\right),
\end{align}
where
$F_{mn}$ is self-dual with
respect to the metric
$ds_\perp^2 = h(r)\left((1+z'(y)^2) dy^2+
y^2 d\Omega_3^2\right)\,$. This metric is conformally
flat. With new coordinates $\tilde y(y)$ such that $ds^2 =
\alpha(\tilde
 y)(d\tilde y^2 + \tilde y^2d\Omega_3^2)$, the instanton
configurations (self-dual field strengths) take the usual form.

To compute $V(Q)$ in general requires knowledge of the embedding
function $z(y)$,  which has been computed by a numerical
shooting technique in \cite{BEEGK}.  Imposing boundary conditions
for the large $y$ behaviour, and requirig
 smooth behaviour in the interior, such that an RG flow 
interpretation is possible, leads to a dependence of the chiral quark
condensate $\langle\bar\psi\psi\rangle$ on the quark mass $m$
and on the temperature.
Depending on the ratio $m/b$,  there are two types of solutions,
which differ  by the topology of the D7-branes.  At large $r$ (or
$y$) the geometry of the D7-branes is $AdS_5 \times S^3$ and
the topology of the $r\rightarrow\infty$ boundary is $S^1 \times
R^3 \times S^3$. For sufficiently large $m/b$, the $S^3$ component
of the D7-geometry contracts to zero size at finite $r > b$. In
this case the D7-brane ``ends'' before reaching the horizon at
$r=b$. However, for sufficiently small $m/b$,  the D7-brane ends
at the horizon, at which point the thermal $S^1$ contracts to zero
size.
Both these types of solutions are plotted in figure 2.
There is a first order phase transition at the critical value of
$m/b \approx 0.92$ where the two types of solution meet
\cite{BEEGK,Kirsch:2004km}.  The $\langle\bar\psi\psi\rangle$ condensate is
non-zero on both sides of this transition, although there is a
discontinuous jump in its value.
\\
\begin{center}
\includegraphics[scale=.6]{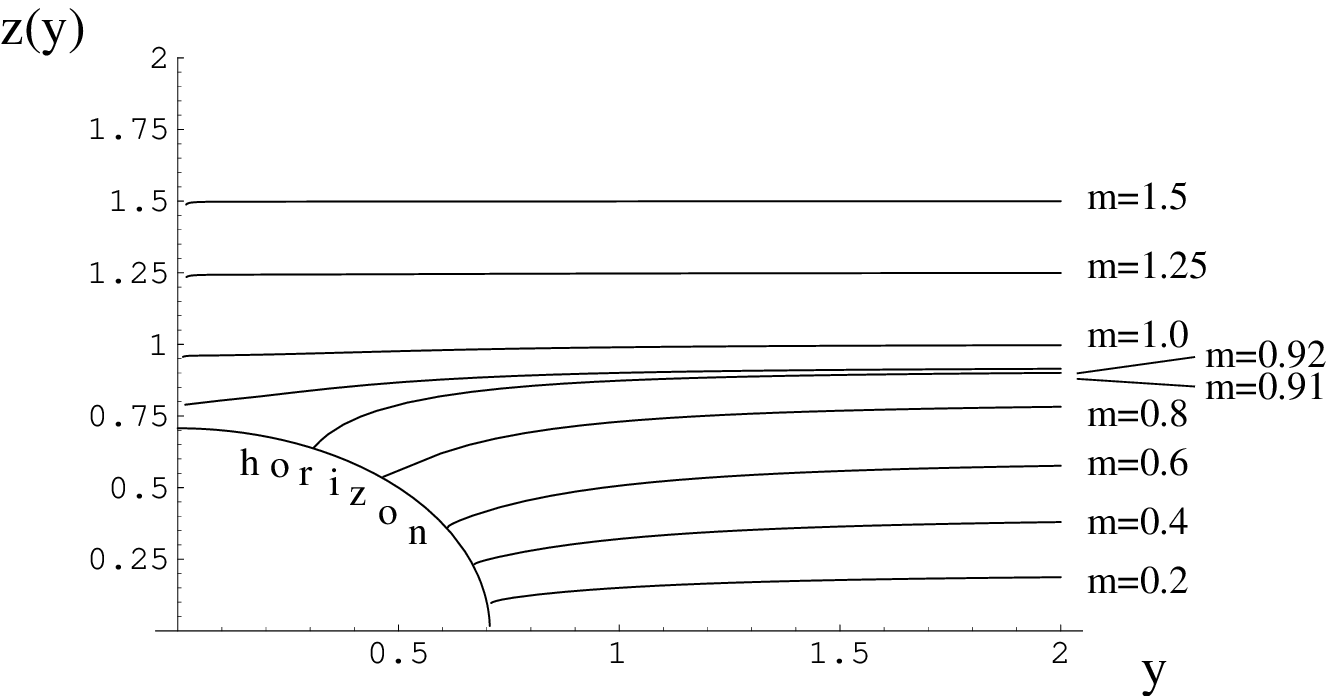}
 {
  \begin{tabular}{r@{ }p{0.7\linewidth}} \small Figure 2. 
    &  \small Brane embeddings in AdS-Schwarzschild for different
  values of the quark mass (with $b=1$).
  \end{tabular}
  }\\[1ex]
\end{center}
This same phase transition is also observed in the Higgs potential as shown in
Figure 3. For $m= \infty$, the potential is flat as in the AdS
case. For smaller and smaller values of $m$, a minimum forms at $Q=0$, until
at a critical value of $m$, the minimum of the potential moves to a finite
value of $Q$. 
\setlength{\unitlength}{1mm}
\begin{center}
\includegraphics[scale=.6]{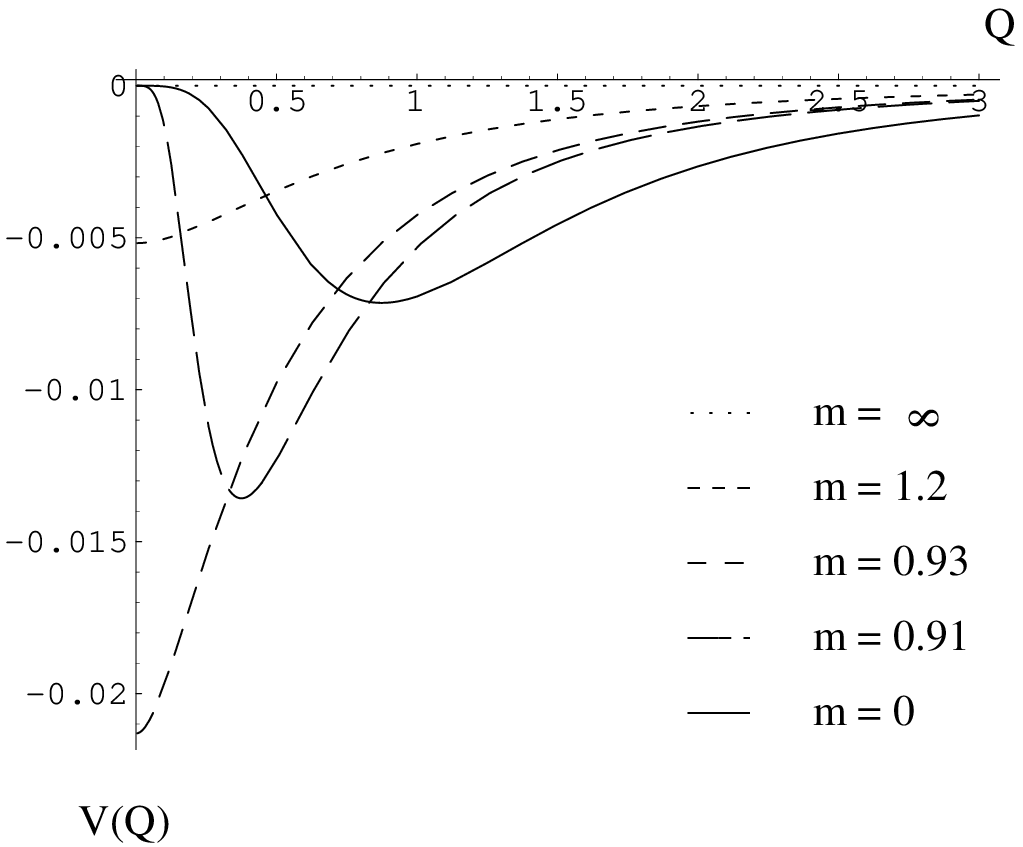}
 {
  \begin{tabular}{r@{ }p{0.7\linewidth}} \small Figure 3. 
    &  \small Potential $V(Q)$ as a function of the instanton size /
Higgs VEV $Q$ for various values of the quark mass $m$. 
  \end{tabular}
  }\\[1ex]

\end{center}
In Figure 4  the Higgs vev  $Q_0$,
 for which the Higgs potential is minimised,
 is plotted versus the
quark mass. This
 clearly displays the first order nature of the phase transition.
$Q_0$ is a suitable order parameter for this transition.

\begin{center}
\includegraphics[scale=.6]{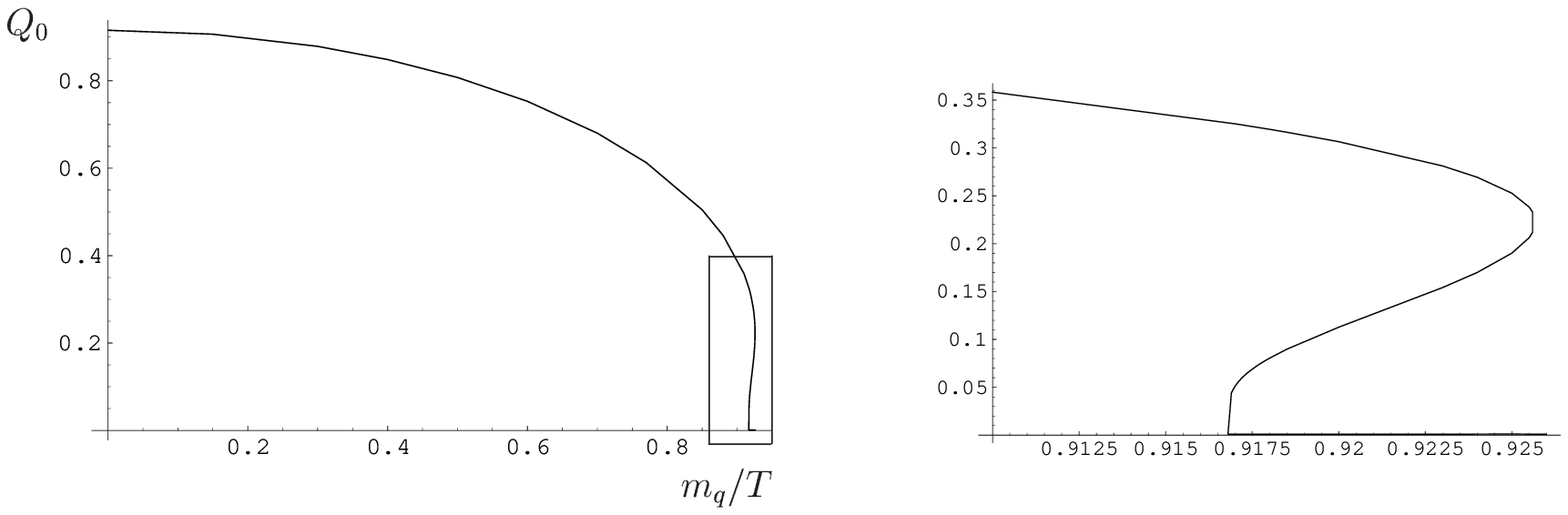}

{
  \begin{tabular}{r@{ }p{0.7\linewidth}} \small Figure 4. 
    &  \small Position of the minimum of the potential $Q_0$ versus the
bare quark mass $m$, zoom of   the critical region.
  \end{tabular}
  }
\end{center}
%
%
%
%

Moreover we have also calculated \cite{AEE} 
the Higgs potential for the background of
Constable and Myers (CM) \cite{Constable} (see also \cite{add2}).
This dilaton-flow geometry 
is asymptotically AdS at large
radius,  but is deformed in the interior 
of the space by an R-chargeless
parameter of dimension four ($b^4$ in what follows).  It is
interpreted as being  dual to ${\cal N}=4$ gauge theory with a non-zero
expectation value for $\tr F^2$. It  was used in
\cite{BEEGK}  to study
chiral symmetry breaking because of its particularly simple form with
a flat six-dimensional plane transverse to the D3 branes. The core of
the geometry is singular\footnote{This singularity may presumably 
be lifted by the D3 branes forming some sort of fuzzy sphere 
in the interior of the space.}.  
In {Einstein frame}, the Constable-Myers geometry is given by
\begin{equation}
ds^2 
=
H^{-1/2} 
K^{\delta/4}
dx_{4}^2 +
H^{1/2} 
K^{(2-\delta)/4} \frac{u^4 - b^4 }{ u^4 }
\sum_{i=1}^6 du_i^2,
\end{equation}
where
\[
K = \left( \frac{ u^4 + b^4 }{ u^4 - b^4}\right),
\qquad H = K^{\delta} - 1,
\qquad \quad \delta =\frac {R^{4}}{2 b^4}, \hspace{1cm} \Delta^2 = 10 - \delta^2 \, ,
\]
\begin{equation}
e^{2 \Phi} = g_s^2 
K^{\Delta}, \hspace{1cm} C_{(4)} =
(g_s \,   H)^{-1} \,
 dt \wedge dx \wedge dy \wedge dz.
\end{equation}

In this geometry, D7
brane probes are repelled by the central singularity,
giving rise to chiral symmetry breaking \cite{BEEGK}. 
This is shown in Figure 5a), where $\sum_{i=1}^4
u_i^2=\rho^2$.

\begin{center}

\includegraphics[scale=.7]{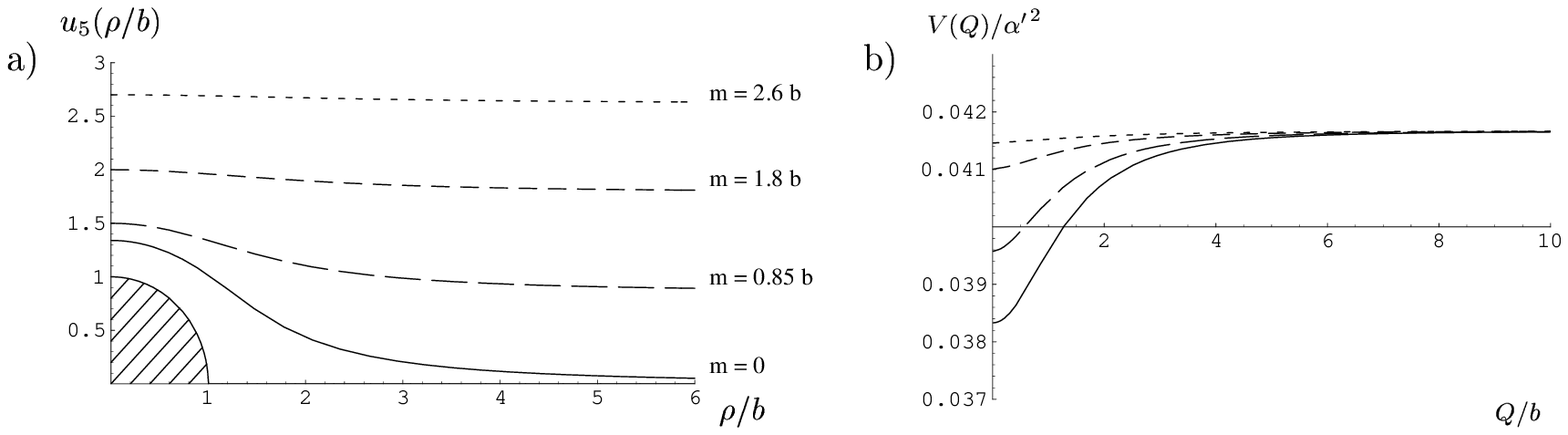}

 {
  \begin{tabular}{r@{ }p{0.7\linewidth}} \small Figure 5. 
    & \small a) Plot of the D7 brane embedding in the Constable-Myers
 geometry for different values of the quark mass $m$.
b) Potential versus  the Higgs vev $Q$ for the values of
$m$ shown in 5a).  
 \end{tabular}
  }\\[1ex]
\end{center}

The Higgs potential (\ref{thepot}) for the Constable-Myers background is
displayed in Figure 5b). We see that although the Constable-Myers
background is expected to be unstable, 
the Higgs potential has a stable minimum at $Q=0$.
Thus it is well-behaved and drives the vev to zero. 
The minimum of the Higgs potential becomes more and more pronounced when the
the quark mass is sent to zero and the
brane is strongly bent, as required for spontaneous chiral symmetry breaking. 
For large $Q$  the potential has the form of a
constant minus a $1/Q^4$ term (remember $Q^2 = \langle \bar{q} q
\rangle$ with $q$ the scalar quarks). 
This behaviour is determined
essentially by dimensional counting since the supersymmetry
breaking operator, tr$F^2$, is dimension four.

Let us provide some intuition for why the brane
configuration disfavours large instantons.
We suggest the
essential reason is that the background metric causes
volume elements to expand for small $\rho$: The 
D7 brane bends away from the singularity in order
to minimize its world-volume. We expect 
that the instanton action  will grow with the size of the instanton in
the region where the brane is strongly bent, preferring
zero size instantons. This argument does not apply in pure
AdS because the four-form term conspires to cancel the
$\sqrt{\det G}$ volume term. However when supersymmetry is broken, this
cancellation no longer works, and the increase in the volume
term is the stronger effect.
This ensures a stable minimum for the Higgs vev.

Finally,  we have also shown
\cite{AEE} that
embedding D7 brane probes into the Yang-Mills* background \cite{James}
leads to a Higgs potential which is bounded.

\vspace{1cm}

{\bf Acknowledgements}

  We are grateful to Christoph Sieg for discussions. The work of R.~A.~and of
Z.~G.~has been supported by the Deutsche Forschungsgemeinschaft (DFG), grants
ER301/1-4 and ER301/2-1. J.~G. acknowledges partial support through the DFG
Graduiertenkolleg 271.


\end{document}